\documentclass[amsmath,amssymb,pra,aps,showpacs,superscriptaddress,twocolumn]{revtex4-2}
\usepackage{amsmath,amsfonts,amssymb,amsthm,graphics,graphicx,epsfig,bbm}
\usepackage[colorlinks=true,citecolor=blue,linkcolor=red,urlcolor=blue]{hyperref}
\usepackage[usenames]{color}
\usepackage{graphicx}
\usepackage{subfigure}
\usepackage{amsmath}
\usepackage{epsfig}
\usepackage{dcolumn}
\usepackage{bm}
\usepackage{color}
\usepackage{epstopdf}
\usepackage{amssymb}
\usepackage{amstext}
\usepackage{latexsym}
\usepackage{hyperref}
\usepackage{amsfonts}
\usepackage{psfrag}
\usepackage{xcolor}
\usepackage[normalem]{ulem}
\usepackage{dsfont}
\usepackage{txfonts}
\usepackage{overpic}
\usepackage{ifthen}
\usepackage{amsthm}

\usepackage{amsmath, amssymb, amsthm}

\newcommand{\ket}[1]{\vert #1 \rangle}

\newcommand{\an}[2]{\ifthenelse{\equal{#1}{}}{\ensuremath{\hat{#1}_{#2}}}{\ensuremath{\hat{#1}^{\protect\phantom{\dagger}}_{#2}}}}

\begin{document}

\title{Quantum fingerprints of self-organization in spin chains coupled to a Kuramoto model}

\author{V. M. Bastidas}
\email{victor.bastidas@ntt.com}
\affiliation{%
Department of Chemistry, Massachusetts Institute of Technology, Cambridge, Massachusetts 02139, USA
} 
\affiliation{%
Physics and Informatics Laboratory, NTT Research, Inc., 940 Stewart Dr., Sunnyvale, California, 94085, USA
} 
\date{\today}

\begin{abstract}

Floquet theory is a widely used framework to describe the dynamics of periodically-driven quantum systems. The usual scenario to describe such kind of systems is to consider the effect of an external control with a definite period in time that can act either locally or globally on the system of interest. However, besides the periodicity, there is no classical correlation or other well-defined structures in the drive. In this work, we consider drives with their own dynamics exhibiting self-organization phenomena and reaching periodic steady states with emergent symmetries. To substantiate our results, we consider two examples of one-dimensional quantum spin chains coupled to a classical Kuramoto model.  First we investigate a Kuramoto model with all-to-all coupling driving a one-dimensional  quantum Ising chain into a time-periodic steady state with an emergent translational symmetry. Next, we consider a Kuramoto model in a Zig-zag lattice driving a XX spin chain. The dynamics of travelling waves in the Kuramoto model trimerizes the lattice, effectively inducing topological behavior that can be exploited to perform topological pumping. Our results can be experimentally implemented in digital and analog near-term quantum devices.

\end{abstract}

\maketitle

\textit{Introduction:}
Self-organization is one of the most intriguing phenomena in nature~\cite{karsenti2008}. Individual units can form complex patterns and structures by following simple dynamical rules~\cite{Cross1993}. These exist due to the delicate balance between nonlinear behavior, energy gain and loss~\cite{jenkins2013}. 

Synchronization is a very relevant example of self-organization in complex dynamical systems~\cite{pikovsky2001,arenas2008,strogatz2018} and even in the quantum regime~\cite{Lee2013, Walter2014, Bastidas2015, Ishibashi2017, Sonar2018, Wachtler2023, Waechtler2024}. The Kuramoto model of coupled phase oscillators~\cite{Acebron2001} is without doubt, one of the cornerstones of theory of synchronization. Phase oscillators host a plethora of patters such as traveling waves~\cite{laing2016}, chimera states~\cite{haugland2021,zakharova2020}, spiral and scroll waves~\cite{shima2004,totz2019synchronization}, among others. They have applications to diverse fields such as electrical networks~\cite{Witthaut2022}, neuronal dynamics~\cite{muller2018}, and chemical oscillators~\cite{mikhailov2006}, just to mention but a few.

Driven quantum manybody systems exhibit a rich behavior~\cite{Eckardt2017}. For example, Floquet systems with a time-periodic external drive exhibit heating phenomena~\cite{DAlessio2014} and they can be used to engineer phases of matter with no counterpart in equilibrium~\cite{oka2019,weitenberg2021,Bastidas2022}. They also exhibit thermal~\cite{Haldar2018,Regnault2016}, manybody localized (MBL)~\cite{Abanin2019}, time crystalline phases~\cite{Zaletel2023,sacha2017time,else2020discrete} and manybody quantum scars~\cite{Haldar2021,Mukherjee2020}. Recent works have explored the effect of quasiperiodic drives on thermalization and other complex dynamics~\cite{Long2022,Pilatowsky2023}. 
The common aspects of these previous works is the use of a classical drive that acts globally or locally on a manybody system. Besides being periodic or quasiperiodic, the drive does not have any dynamics or further structure. Therefore a natural question becomes: what are the quantum signatures of a drive with its own dynamics and which kind of behaviors it can induce on a manybody system.

In our work, we consider a manybody system under the effect of a drive whose dynamics is governed by the Kuramoto model~\cite{Acebron2001}. More specifically, we use the Kuramoto model to drive a one-dimensional quantum spin chain~\cite{mikeska2004}. 
We numerically solve the classical equations for the Kuramoto model for a given network topology and use the solution as a local drive for a quantum spin chain. Notably, at initial times there are no symmetries present in the quantum system as the phase oscillators are initialized in random initial conditions. However, during the dynamics, the Kuramoto model exhibits pattern formation and we investigate how it influences the quantum dynamics of the chain.

We focus on two different scenarios. First we couple a quantum Ising chain in a transverse field~\cite{Dziarmaga2005} to a Kuramoto model with all-to-all coupling. The latter exhibits a transition to synchronized motion after a transient time. This effectively drives the system from a structureless initial state with no symmetries before the transient time, to a system with translational invariance in space and time, which are emergent symmetries during the dynamics. We also consider a Kuramoto model in a zig-zag lattice with unidirectional coupling acting as a drive for a quantum XX model~\cite{Riddell2019}. After a transient time, the classical drive forms a periodic traveling wave pattern that breaks time-reversal symmetry of the quantum model. This induces an emergent topological behavior that can be exploited to perform topological pumping of edge excitations in the chain.

The field of driven quantum systems beyond Floquet and quasiperiodic drives is widely unexplored. A recent work has predicted that quasiperiodic drives induce complete Hilbert-space ergodicity~\cite{Pilatowsky2023}. Our main contribution is to show that the dynamics of a self-organized drive induce rich behavior on our quantum spin chain and is responsible for emergent symmetries. In future works it would be interesting to investigate quantum signatures of pattern formation in non-integrable models and to explore the phases that appear in this scenario.

\begin{figure}
	\includegraphics[width=0.40\textwidth]{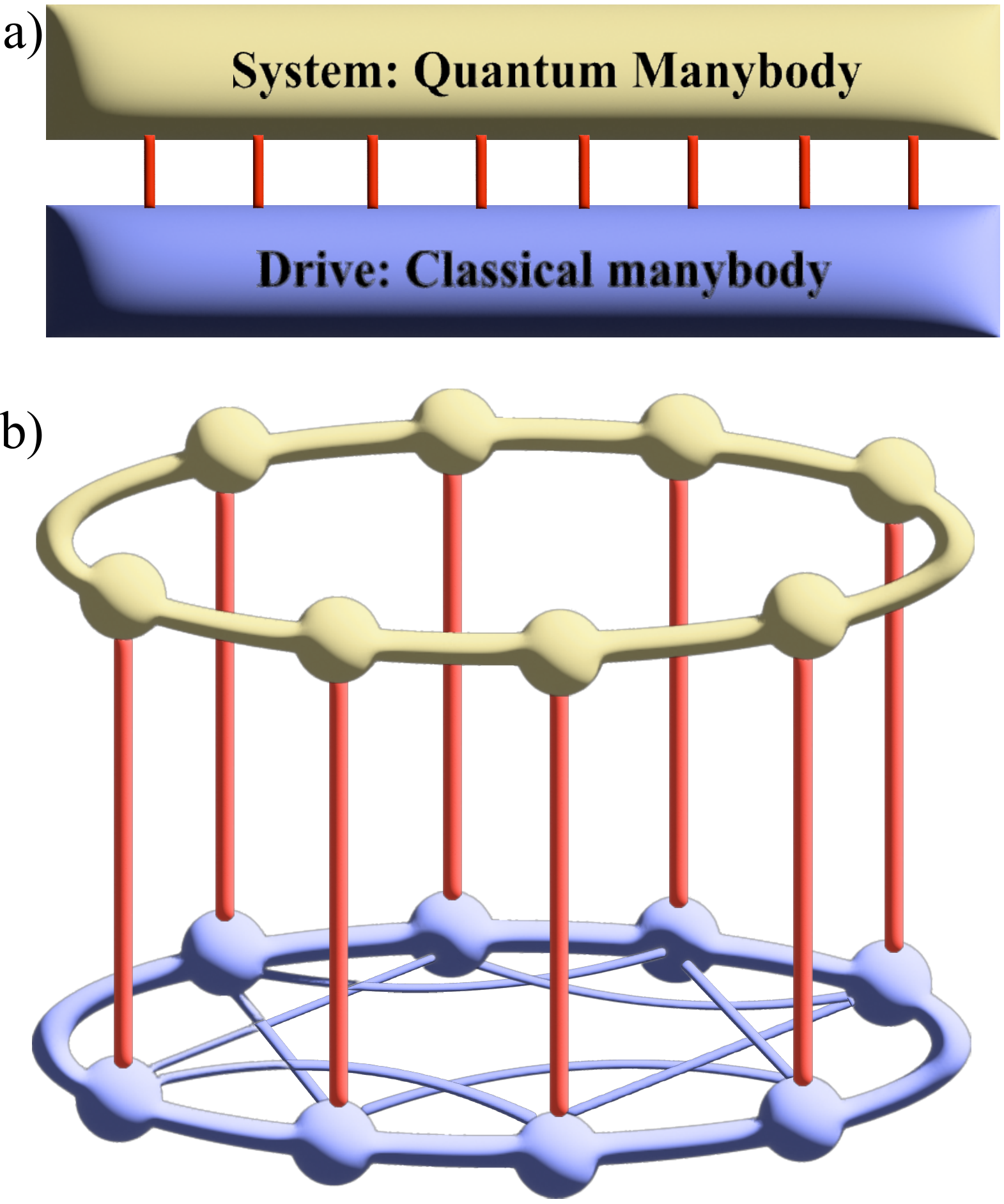}
	\caption{a) A quantum manybody system dynamics driven by a classical manybody system. The classical drive acts locally on the quantum manybody system. b) Illustrates a manybody lattice system with short range interactions coupled to a network of with arbitrary connectivity. }
	\label{Fig1}
\end{figure}

\textit{XY spin chain in a transverse field coupled to a Kuramoto model.} In our work we focus on a time dependent XY quantum spin chain in a transverse field~\cite{mikeska2004,sachdev_2011,McKenzie1996}
\begin{align}
          \label{eq:XYHamitonian}
\hat{H}(t)&=\hbar \sum^{N}_{i=1}g_i(t) \sigma^z_j-\hbar \sum^{N}_{i=1}(J_x \sigma^x_i\sigma^x_{i+1}+J_y\sigma^y_i\sigma^y_{i+1})
\ ,
\end{align}
where $\sigma^x_j,\sigma^y_j,\sigma^z_j$ are Pauli operators acting on the $j$-th qubit. We consider a local time dependent modulation $g_i(t)=G\cos[\theta_i(t)] $ of the transverse field with amplitude $G$, while $J_x$ and  $J_y$ are the anisotropic interaction strengths in $x$ and $y$ directions, respectively. The time-dependent phases $\{\theta_i(t)\}$ determining the local transverse field are obtained by solving the classical equation for the Kuramoto model of coupled phase oscillators~\cite{Acebron2001}
\begin{align}
          \label{eq:KuramotoModel}
\dot{\theta_i}(t)&=\omega_i+\sum^N_{j=1}K_{i,j}\sin[\theta_j(t)-\theta_i(t)+\beta_j]
\ ,
\end{align}
where 
$\omega_i$ are local frequencies drawn from a normal distribution while $\beta_j$ is a phase delay. The initial phases  $\theta_i(0)\in[-\pi,\pi]$ are chosen randomly from a uniform distribution. The Kuramoto model is a paradigmatic model in nonlinear dynamics that shows the transition to synchronized behavior. In our work, its solution acts as a classical drive for the quantum spin chain, as we illustrate in Fig.~\ref{Fig1}. One important point here is to note here that while the XY spin chain is a Hamiltonian system, the Kuramoto model does not have, in general, a Hamiltonian representation (see appendix~\ref{AppendixA} for further details)

\begin{figure*}
	\includegraphics[width=0.95\textwidth]{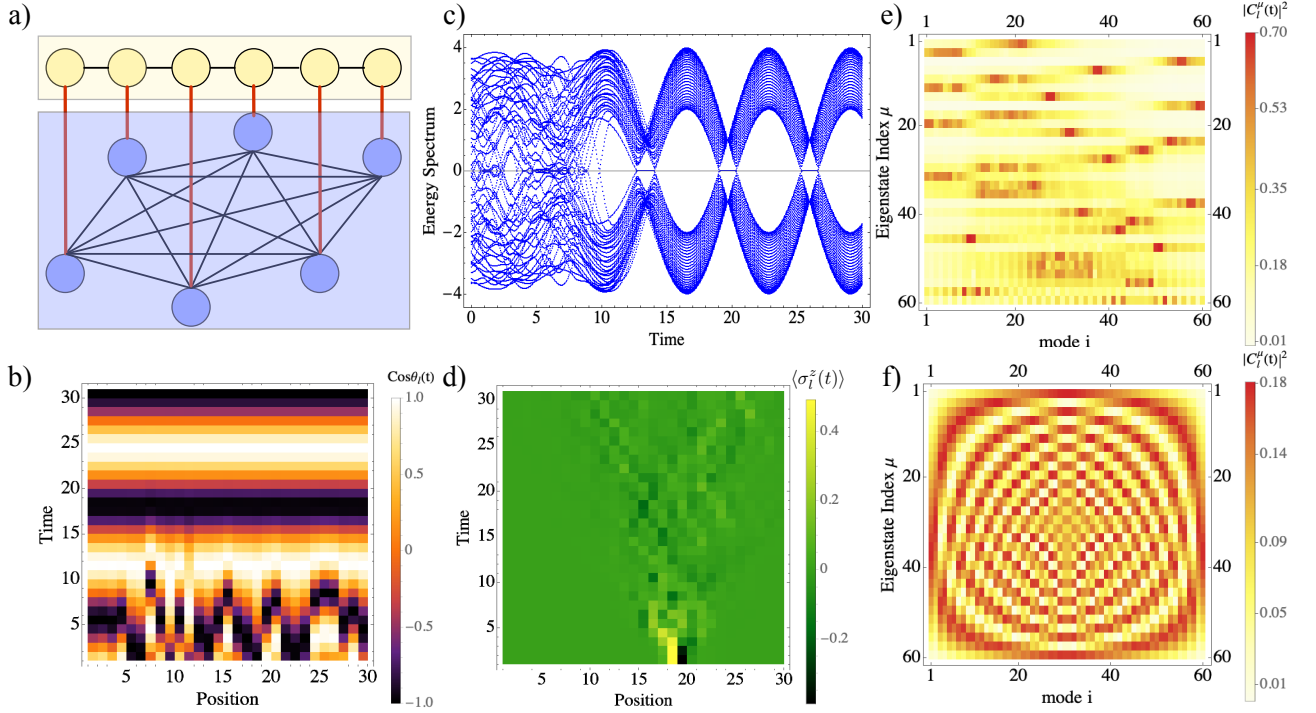}
	\caption{a) Illustrates a one-dimensional quantum Ising chain with $N=30$ sites that is coupled to a network of $N=30$ phase oscillators with all-to-all coupling described by the Kuramoto model. b) Shows the time evolution of the transverse local fields $g_i(t)=G\cos[\theta_i(t)] $ and the emergence of synchronized motion. c) Depicts the instantaneous spectrum and the formation of two energy bands periodically oscillating in time. d) Shows the time evolution of a localized energy eigenstate of the Hamiltonian $\hat{H}(0)$. e) and f) Depict two snapshots of the eingenstates before ($J t=5$) and after ($J t=23$) synchronization. We have chosen parameters $G=3J$, $\tilde{K}=0.5J$, and $\omega=0.5J$.}
	\label{Fig2}
\end{figure*}

It is instructive to represent our spin chain in terms of fermions. To do this, we use the Jordan Wigner transformation~\cite{jordan1928}
\begin{align}
          \label{eq:JordanWigner}
 \hat{f}_j=\frac{1}{2}\left(\prod_{l=1}^{j-1}\sigma^z_l\right)(\sigma^x_j+\mathrm{i}\sigma^y_j)
 \ ,
\end{align}
where $\sigma^z_l=1-2\hat{f}^{\dagger}_j\hat{f}_j$  and $\hat{f}^{\dagger}_j$ and $\hat{f}_j$ are the fermionic creation and annihilation operators satisfying the anticommutation relations $\{\hat{f}_i,\hat{f}^{\dagger}_j\}=\delta_{i,j}$ and $\{\hat{f}_i,\hat{f}_j\}=\{\hat{f}^{\dagger}_i,\hat{f}^{\dagger}_j\}=0$. 

After applying the Jordan Wigner transformation to the Ising model in Eq.~\eqref{eq:XYHamitonian}, we obtain the fermionic quadratic Hamiltonian~\cite{Dziarmaga2005}
\begin{align}
          \label{eq:IsingHamitonianFermionSI}
\hat{H}(t)&=-\hbar \sum^{N}_{j=1} g_j(t) (1-2\hat{f}^{\dagger}_j\hat{f}_j)- \hbar \sum^{N-1}_{j=1}(J_x+J_y)(\hat{f}^{\dagger}_j\hat{f}_{j+1}+\hat{f}^{\dagger}_{j+1}\hat{f}_{j})
\nonumber\\&
 -\hbar \sum^{N-1}_{j=1}(J_x-J_y)(\hat{f}^{\dagger}_j\hat{f}^{\dagger}_{j+1}+\hat{f}_{j+1}\hat{f}_{j})
\ ,
\end{align}
which can also be written as 
\begin{align}
          \label{eq:IsingHamitonianFermionMajorana}
\hat{H}(t)= \mathrm{i} \hbar\sum^{N}_{j=1} g_j(t) \hat{a}_{2j-1}\hat{a}_{2j}+  \mathrm{i}  \hbar \sum^{N-1}_{j=1}(J_x\hat{a}_{2j}\hat{a}_{2j+1}
-J_y\hat{a}_{2j-1}\hat{a}_{2j+2})
\end{align}
in terms of Majorana fermions $\hat{a}_{2j-1}=\hat{f}^{\dagger}_j+\hat{f}_j$ and $\hat{a}_{2j}=\mathrm{i}(\hat{f}^{\dagger}_j-\hat{f}_j)$~\cite{kitaev2001unpaired,Benito2014} satisfying the anticommutation relations $\{\hat{a}_m,\hat{a}_n\}=\delta_{m,n}$ given that $\hat{a}_n=\hat{a}^{\dagger}_n$.

To investigate quantum fingerprints of self-organization, we numerically solve the Heisenberg equations of motion for the fermionic operators in several situations. This enables us to explore how symmetries emerge in our quantum system as a consequence of self-organization in the Kuramoto model.

\textit{Emergent symmetries in the Ising model:}
In this section, we will consider a concrete example of a one-dimensional quantum Ising model in a transverse field~\cite{sachdev_2011,Dziarmaga2005}, which is obtained from Eq.~\eqref{eq:XYHamitonian} by setting $J_y=0$. For simplicity, we consider a spin chain with open boundary conditions as depicted in Fig.~\ref{Fig2}~a). We also consider a Kuramoto model of phase oscillators $\dot{\theta_i}(t)=\omega+\tilde{K}\sum^N_{j=1}\sin[\theta_j(t)-\theta_i(t)]$ in a network with all-to-all coupling $K_{i,j}=\tilde{K}$, uniform frequencies $\omega_i=\omega$ and with no phase delay $\beta_j=0$ [see Fig.~\ref{Fig2}~a)]. 

As a first step to investigate the dynamics, we numerically solve the equations of motion for the Kuramoto model. As we choose random initial conditions, the local transverse field $g_i(t)=G\cos[\theta_i(t)] $ breaks the translational invariance of the quantum Ising chain as we show in Fig.~\ref{Fig2}~b). In this figure one can see that the system synchronizes after a transient time and the phases $\{\theta_i(t)\}$ for each $i$ become identical and oscillate in time with a period $T$.  

Next we will discuss how the onset of synchronization influences the quantum dynamics of the spin chain. To study the dynamics it is convenient to work with the Hamiltonian~\eqref{eq:IsingHamitonianFermionMajorana} in terms of Majorana fermions~\cite{kitaev2001unpaired,Benito2014} by setting $J_y=0$. From this we can derive the Heisenberg equations of motion 
\begin{align}
          \label{eq:HeisenbergEqMajorana}
\mathrm{i}\frac{d \hat{a}_{2i-1}}{dt}&=\mathrm{i}g_i(t)\hat{a}_{2i}-\mathrm{i}J_x\hat{a}_{2(i-1)} \ ,
\nonumber \\
\mathrm{i}\frac{d \hat{a}_{2i}}{dt}&=-\mathrm{i}g_i(t)\hat{a}_{2i-1}+\mathrm{i}J_x \hat{a}_{2(i+1)-1}
\ .
\end{align}
In the appendices~\ref{AppendixB}~and~\ref{AppendixC} we briefly discuss the continuous limit of this theory and its relation to the Dirac equation. Before synchronization takes place in the Kuramoto model, there is no translational invariance in space and time as the phases are random and there is no periodicity in time. However, when the phase oscillators synchronize, there are  emergent translational symmetries in space and time. Effectively, the system becomes a Floquet system with time-periodic Hamiltonian $\hat{H}(t+T)=\hat{H}(t)$ and the quasimomentum $k$ becomes a good quantum number because the system is translationally invariant with $\theta_i(t)=\theta_j(t)=\Omega t+\psi_0$ for all $i\neq j$. In this regime, the system is described by the Hamiltonian in the quasimomentum space~\cite{Dziarmaga2005}
\begin{align}
          \label{eq:IsingMomentum}
                 \hat{H}(t)&=-2\sum_k[g(t)+J\cos k](\hat{F}_k^{\dagger}\hat{F}_k-\hat{F}_{-k}
                 \hat{F}_{-k}^{\dagger})
                 \nonumber \\
                 &+2J_x\sum_k\sin k(\hat{F}_k^{\dagger}\hat{F}^{\dagger}_{-k}+\hat{F}_{-k}\hat{F}_{k})
\ ,
\end{align}
 where $g(t)=G\cos(\Omega t+\psi_0)$ and $\hat{F}_k$ are fermionic operators. In this example, discrete translational invariance in space and time are emergent symmetries in our quantum model.

Let us analyze the instantaneous spectrum shown in Fig.~\ref{Fig2}~c). In this figure one clearly see that the spectrum does not have much structure before synchronization takes place. When the system synchronizes, there are two well-defined energy bands that exhibits periodic oscillations. This can be attributed to the existence of translational invariance in our system in space and time. Figure~\ref{Fig2}~d) shows the dynamics of the expectation value $\langle \sigma^z_i(t)\rangle$. The initial state was chosen to be a localized energy eigenstate of the Hamiltonian $\hat{H}(0)$. As expected, before classical synchronization takes place, the state remains localized. After the phase oscillators synchronize, there is a ballistic spreading along the chain. The dynamics of the Kuramoto model also influences the behavior of the eigenstates before and after synchronization takes place as we depict in  Figs.~\ref{Fig2}~e)~and~f), respectively. There one can see that when the phase oscillators are synchronized, there is a structure of the eigenstates that show oscillations in space associated to the quasimomentum.

\begin{figure*}
	\includegraphics[width=0.95\textwidth]{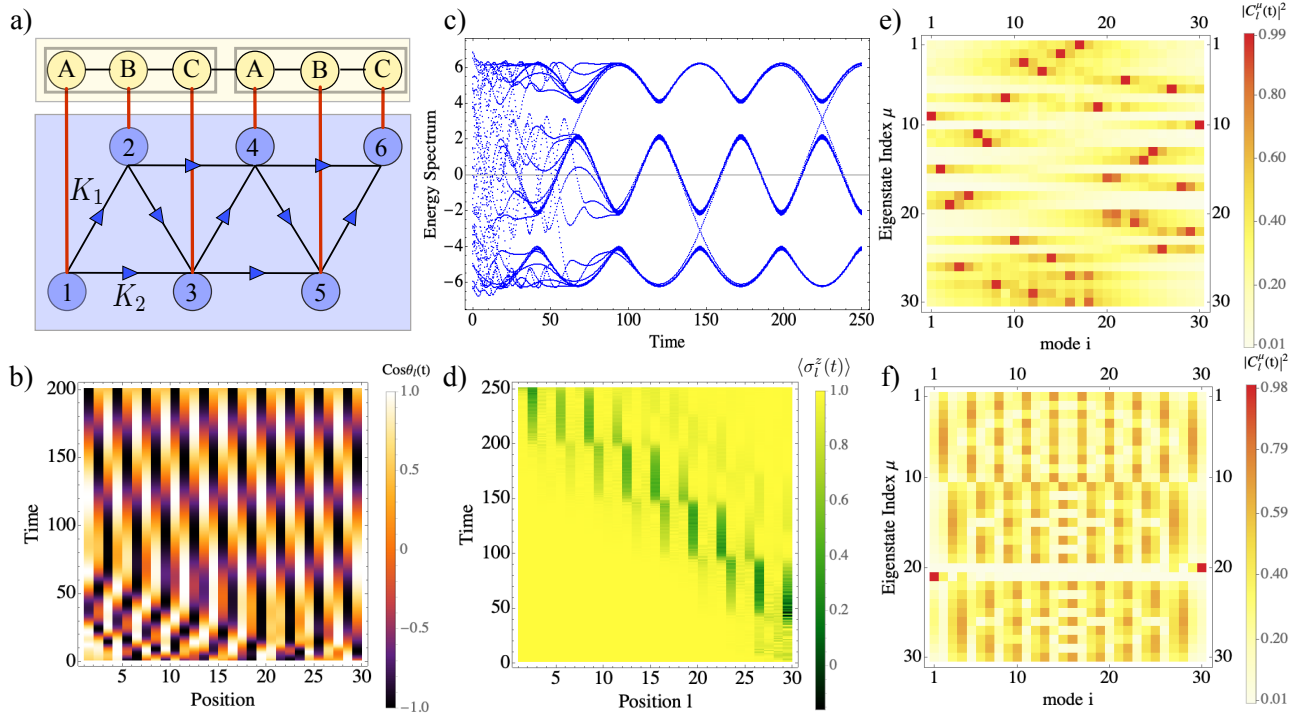}
	\caption{a) Illustrates a one-dimensional quantum Ising chain with $N=30$ sites that is coupled to a Kuramoto model in a zig-zag lattice with unidirectional coupling. b) Shows the time evolution of the transverse local fields $g_i(t)=G\cos[\theta_i(t)] $ and the emergence of travelling waves after the system synchronizes. c) Depicts the instantaneous spectrum and the emergence of three energy bands periodically oscillating in time. d) Shows the time evolution of a localized energy eigenstate of the Hamiltonian $\hat{H}(0)$ and how it is pumped along the lattice. e) and f) Depict two snapshots of the eingenstates before ($\omega t=50$) and after ($\omega t=150$) synchronization. We have chosen parameters $G=3J$, $K_1=0.2J$, $K_2=0.1J$, and $\omega=0.04J$}
	\label{Fig3}
\end{figure*}

\textit{Emergent symmetry breaking and topological pumping in the XX model:}
So far we have discussed how the dynamics classical Kuramoto model is responsible for the emergent symmetries of our quantum spin chain. Next, we will consider a different situation where the classical phase oscillators restore spatial symmetries while breaking time reversal symmetry in a XX model [$J_x=J_y=J$ in Eq.~\eqref{eq:XYHamitonian}] with open boundary conditions[see Fig.~\ref{Fig3}~a)]. Here we consider a Kuramoto model $\dot{\theta_i}(t)=\omega+\sum^{2}_{r=1}K_r\sum^{N-1}_{j=1}\sin[\theta_{i+r}(t)-\theta_i(t)-2r\pi/3]$ with unidirectional coupling and phase delay $\beta_r=-2r\pi/3$. As we depict in Fig.~\ref{Fig3}, the oscillators are arranged in a zig-zag geometry with couplings $K_1$ and $K_2$.

We numerically solve the equations of motion for the phase oscillators with random initial conditions. Similarly to the Ising model, at transient times before synchronization appears, the phases are random and the dynamics breaks the translational invariance of the quantum model in space. After synchronization, the Kuramoto dynamics exhibits periodic traveling waves~\cite{chow1998traveling} with a very low frequency in time as we show in Fig.~\ref{Fig3}~b). This effectively generates a pattern of phases in space that moves with a preferred direction that breaks time reversal in the quantum model.

As the XX model preserves the total number of fermions, to explore its dynamics we need to solve the Heisenberg equations of motion for the fermionic operators
\begin{align}
          \label{eq:HeisenbergEqMajorana}
\mathrm{i}\frac{d \hat{f}_{i}}{dt}&=-2g_i(t)\hat{f}_{i}-2J(\hat{f}_{i+1}+ \hat{f}_{i-1})
\ .
\end{align}
The solution of the Kuramoto model acts as a local transverse field for the XX chain. Thus, by solving the equation of motion we can explore quantum signatures of the traveling waves. Fig.~\ref{Fig3}~c) shows the instantaneous spectrum. At short times, it does not exhibit any structure as the phase oscillators are unsynchronized. When synchronization takes place, the traveling wave pattern effectively trimerize the spin chain~\cite{Haug2020} because the transverse field reads $g_i(t)=G\cos[\Omega t+2\pi i/3+\phi_0] $. For this reason we see three well defined bands associated to the traveling wave. This behavior resembles a situation encountered in the theory of the quantum Hall effect and it is intimately related to topological pumping~\cite{Thouless1983,niu1984quantised}. The traveling wave acts as an effective adiabatic drive breaking time reversal symmetry, which allows us to adiabatically pump charges in the system as we show in Fig.~\ref{Fig3}~d). Before synchronization, the eigenstates are mostly localized due to the random transverse fields. However, when the system is synchronized, the eigenstates exhibit a spatial structure due to the trimerization of the lattice.

One of the key points of this example is that while the Kuramoto drive is responsible for the emergence of a well-defined trimer structure, it also breaks the time-reversal symmetry giving rise to a topological transport process.

\textit{Conclusions:}
In summary, we have shown quantum signatures of a self-organization by considering a Kuramoto model coupled to a spin chain. In particular we have demonstrated how symmetries can emerge due to pattern formation in our drive. Of course, we are not restricted to one-dimensional spin chains. In fact, our work is also intimately related to quantum simulation of electronic structure in quantum computers driven by the motion of nuclei within the framework of the Born-Oppenheimer approximation~\cite{Born1927}. In this context, one needs to solve the equation of motion for the classical nuclei, and the solution modulates the one- and two-particle integrals of the electronic structure Hamiltonian (see appendix ~\ref{AppendixD} for further details).

In future works, it would be interesting to study two-or three dimensional lattices driven by Kuramoto or coupled Stuart Landau Oscillators to generate more complex dynamics.  Recent works have shown that even a quasiperiodic drive can induce complex dynamics leading to complete Hilbert space ergodicity~\cite{Pilatowsky2023}. Our approach can open a new avenue to study complex dynamics in manybody systems. For example,
it would be interesting to explore which phases are generated under the effect of self-organized drives.

{\it{Acknowledgments.---} }
The author thanks D. B. Sato for valuable discussions, and NTT Research
Inc.  for their support during this project.

\appendix

\section{A quantum XY model coupled to a network of Stuart Landau oscillators
\label{AppendixA}}
Similarly to Ref.~\cite{Bastidas2015}, we consider a quantum network consisting of a ring of $N$ coupled Van der Pol oscillators.
Such a network can be described by the master equation for the density matrix $\rho(t)$~\cite{Wachtler2023}
\begin{align}
      \label{MasterEqNetwork}
            \dot{\rho}&=-\frac{\mathrm{i}}{\hbar}[\hat{H}_{\text{XYSL}},\rho]+2\sum^{N}_{l=1}\left[\kappa_{1}\mathcal{D}(\hat{A}_{l}^{\dagger})+\kappa_{2}\mathcal{D}(\hat{A}_{l}^{2})\right]\rho
      \ ,
\end{align}
where $\hat{A}^{\dagger}_{l}$ and $\hat{A}_{l}$ are creation and annihilation operators of bosonic particles and  $\mathcal{D}(\hat{O})\rho=\hat{O}\rho\hat{O}^{\dagger}-\frac{1}{2}(\hat{O}^{\dagger}
\hat{O}\rho+\rho\hat{O}^{\dagger}
\hat{O})$ describes dissipative processes with rates $\kappa_1,\kappa_{2}>0$.

The coherent part of the Hamiltonian describes the coupling between the quantum oscillators and the Ising chain and is given by

\begin{align}
      \label{MasterEqNetwork}
            \hat{H}_{\text{XYSL}}&=
-\hbar \sum^{N}_{i=1}(J_x \sigma^x_i\sigma^x_{i+1}+J_y\sigma^y_i\sigma^y_{i+1})+ \hbar \sum^{N}_{i=1}\omega_j \hat{A}^{\dagger}_{i}\hat{A}_{i}
\nonumber \\
+& \hbar \sum^{N}_{i=1}G(\hat{A}_{i}+\hat{A}^{\dagger}_{i}) \sigma^z_i+ \sum_{i,j=1}^{N}K_{i,j}(e^{\mathrm{i}\beta_j}\hat{A}^{\dagger}_{i}\hat{A}_{j}+e^{-\mathrm{i}\beta_j}\hat{A}_{i}\hat{A}^{\dagger}_{j})
      \ .
\end{align}

Next, we will formalize a semiclassical expansion of the bosonic modes  $\hat{A}_{l}(t)=\alpha_{l}(t)+\hat{a}_{l}$, where $\bm{\alpha}(t)=[\alpha_1(t),\ldots,\alpha_{N}(t)]$ is the semiclassical trajectory, and $\hat{\bm{a}}=(\hat{a}_{1},\ldots,\hat{a}_{N})$ describes the quantum fluctuations.
We will work in the semiclassical regime where the magnitude of $\alpha_{i}(t)$ is much larger than the quantum fluctuations described by $\hat{a}_{i}$. 

Let us consider the bosonic displacement operator $\hat{D}\left[\bm{\alpha}(t)\right]=\exp\left[\bm{\alpha}(t) \cdot \hat{\bm{a}}^{\dagger}-\bm{\alpha}^{*}(t)\cdot\hat{\bm{a}}\right]$, defining the time-dependent coherent states $\ket{\alpha_{i}(t)}=\hat{D}\left[\bm{\alpha}(t)\right]\ket{0_{i}}$, where $\ket{0_{i}}$ is the vacuum state of the $i$-th oscillator.

Following the same method as in Ref.~\cite{Bastidas2015}, we will consider a expansion of the density operator $\rho_{\bm{\alpha}}(t)=\hat{D}^{\dagger}\left[\bm{\alpha}(t)\right]\rho(t)\hat{D}\left[\bm{\alpha}(t)\right]$ in a co-moving frame defined by the semiclassical trajectory $\bm{\alpha}(t)$. This allows us to obtain a master equation for the  density operator  
\begin{equation}
      \label{SemiclassMasterNetwork}
        \dot{\rho}_{\bm{\alpha}}\approx-\frac{\mathrm{i}}{\hbar}[\hat{H}_{\bm{\alpha}},\rho_{\bm{\alpha}}]+2\sum^{N}_{i=1}\left[\kappa_{1}\mathcal{D}(\hat{a}_{i}^{\dagger})
        +4\kappa_{2}|\alpha_{i}|^2\mathcal{D}(\hat{a}_{i})\right]
      \ .
\end{equation}
In addition, the coherent dynamics of the fluctuations are governed by
a Hamiltonian driven by the mean field, as follows
\begin{align}
      \label{EffQuadHamNet}
            \frac{H_{\bm{\alpha}}}{\hbar}&=\sum^{N}_{i=1}G(\hat{a}_{i}+\hat{a}^{\dagger}_{i}+\alpha_{i}+\alpha_{i}^{*}) \sigma^z_i- \sum^{N}_{i=1}(J_x \sigma^x_i\sigma^x_{i+1}+J_y\sigma^y_i\sigma^y_{i+1})
            \nonumber \\
            &+ \sum^{N}_{i=1}\omega_i \hat{a}^{\dagger}_{i}\hat{a}_{i}+\mathrm{i}\sum^{N}_{i=1}\kappa_{2}(\alpha_{i}^{*})^{2}\hat{a}_{i}^{2}
            +\sum_{i,j=1}^{N}K_{i,j}e^{\mathrm{i}\beta_j}\hat{a}^{\dagger}_{i}\hat{a}_{j}+\text{H.c}
       \ .
\end{align}
In turn,  the mean field defining the co-moving frame satisfies the Stuart-Landau equation
\begin{equation}
      \label{QuantEqMotionNetwork}
            \dot{\alpha}_{i}(t)=-\mathrm{i}\omega_i\alpha_{i}(t)+\alpha_{i}(t)(\kappa_{1}-2\kappa_{2}|\alpha_{i}(t)|^2)-\mathrm{i}\sum_{i,j=1}^{N}K_{i,j}e^{\mathrm{i}\beta_j}\alpha_{j}(t).
\end{equation}
In Eq.~\eqref{EffQuadHamNet}, the coupling between the spin chain and the oscillators is via the quantum fluctuations that are smaller in magnitude in comparison to the mean field $|\hat{a}_{i}|\ll |\alpha_i|$. For this reason, we can neglect the coupling between the spin chain and the bosonic fluctuations to obtain an effective theory for the spin chain

\begin{align}
      \label{EffQuadHamNet}
            \hat{H}(t)= \hbar \sum^{N}_{i=1}G(\alpha_{i}+\alpha_{i}^{*}) \sigma^z_i-\hbar \sum^{N}_{i=1}(J_x \sigma^x_i\sigma^x_{i+1}+J_y\sigma^y_i\sigma^y_{i+1})
       \  .
\end{align}
Here the motion of the classical oscillator drives the quantum dynamics of the spin chain. Further, we can use the polar decomposition $\alpha_{i}(t)=r_i(t)e^{\mathrm{i}\theta_i(t)}$ in terms of amplitude $r_i(t)$ and phase $\theta_i(t)$. The equations of motion for these variables
\begin{align}
      \label{Eq;AplitudePhase}
            \dot{r}_{i}&=r_i(\kappa_{1}-2\kappa_{2}r_{i}^2)-r_i\sum_{i,j=1}^{N}K_{i,j}\sin(\theta_j-\theta_i+\beta_j)
            \nonumber \\
            \dot{\theta}_{i}&=-\omega_i-\sum_{i,j=1}^{N}K_{i,j}\cos(\theta_j-\theta_i+\beta_j)
            \ .
\end{align}
The main difference between the Start Landau Oscillator and the Kuramoto model in the main text is that in the former, the amplitude dynamics is coupled to the phase. Here we analyzed the Stuart-Landau oscillator, as it can be derived from a master equation in the semiclassical limit. To the best of our knowledge, there is not a clear way to derive the Kuramoto model as a semiclassical limit of a quantum model.

\section{Continuous limit of the model: Dirac equation with time dependent mass
\label{AppendixB}}
The first step to systematically define the continuous limit is to consider a lattice constant $b$ defining the spatial coordinate $x_j=jb$ associated to the $j$-th lattice site~\cite{fradkin2021quantum}. We also fermionic field operators depending on this coordinate as $\hat{\chi}_1(x_j)=\hat{a}_{2i-1}/\sqrt{b}$ and $\hat{\chi}_2(x_j)=\hat{a}_{2i}/\sqrt{b}$. This will ensure the right anticommutation relations of the field operators in the continuous limit. Next, let us evaluate the approximate form of the field operators under spatial translations~\cite{sachdev_2011}, as follows

\begin{align}
          \label{eq:DisplacementMajorana}
\hat{\chi}_1(x_j\pm b)&=\hat{a}_{2(j\pm1)-1}/\sqrt{b}\approx \hat{\chi}_1(x_j)\pm b \partial_x \hat{\chi}_1(x_j) \ ,
\nonumber \\
\hat{\chi}_2(x_j\pm b)&=\hat{a}_{2(j\pm1)}/\sqrt{b}\approx \hat{\chi}_2(x_j)\pm b\partial _x\hat{\chi}_2(x_j)
\ .
\end{align}
After inserting these relations in Hamiltonian Eq.~\eqref{eq:IsingHamitonianFermionMajorana} we can derive the Hamiltonian in the continuous limit 
\begin{align}
          \label{eq:SIIsingHamitonianFermionMajorana}
\hat{H}(t)= \mathrm{i} \hbar\int \hat{\chi}_2(x)\left[-\Delta(x,t)+c\partial_x\right]\hat{\chi}_1(x)+E_0(t)
\ ,
\end{align}
where $\Delta(x,t)=g(x,t)-J_x-J_y$, $c=(J_x-J_y)b$ and $E_0(t)$ is a time dependent energy.
Correspondingly, we can also write the continuous limit of the equations of motion in Eq.~\eqref{eq:HeisenbergEqMajorana}
\begin{align}
          \label{eq:HeisenbergEqMajorana}
\mathrm{i}\partial_t \hat{\chi}_1(x)&=\mathrm{i}\left[\Delta(x,t)+c \partial_x\right]\hat{\chi}_2(x) \ ,
\nonumber \\
\mathrm{i}\partial_t \hat{\chi}_2(x)&=-\mathrm{i}\left[\Delta(x,t)-c \partial_x\right]\hat{\chi}_1(x)
\ .
\end{align}
By considering the Dirac matrices $\gamma^0=-\sigma^y$ and $\gamma^1=\mathrm{i}\sigma^z$ one can write the equations of motion as a Dirac equation $\mathrm{i}[\gamma^{\mu}\partial_{\mu}-m(\boldsymbol{x})]\hat{\boldsymbol{\chi}}(\boldsymbol{x})$ in $1+1$ space time with coordinates $\boldsymbol{x}=(X_0,X_1)$ (see Ref.~\cite{fradkin2021quantum}). Here $X_0=ct$ and $X_1=x$ denote the temporal and spatial coordinates, respectively. We also have defined the mass term $m(\boldsymbol{x})=\Delta(x,t)/c$ and the Dirac spinor field $\hat{\boldsymbol{\chi}}^{\dagger}(\boldsymbol{x})=[ \hat{\chi}_1(\boldsymbol{x}), \hat{\chi}_2(\boldsymbol{x})]$.

In the effective field theory, the mass term is also modulated by the dynamics of the Kuramoto model Eq.~\eqref{eq:KuramotoModel} in the continuous limit
\begin{align}
          \label{eq:KuramotoModelField}
          \partial_t\theta(x,t)&=\omega(x)+\int K(x-y)\sin[\theta(x,t)-\theta(y,t)]
\ .
\end{align}
Thus, as $m(\boldsymbol{x})=\Delta(x,t)/c=G\cos[\theta(x,t)]/c$, the dynamics of the Kuramoto model modulates the mass of the Dirac equation in space and time

\section{Dirac fermion coupled to a gravitational field
\label{AppendixC}}
Another example is the relativistic dynamics of electrons coupled to a semiclassical gravitational field. In this case, the time-dependence of the external field enters as a time-dependent mass of the Dirac fermion, governed by the hamiltonian~\cite{koke2016dirac}
\begin{align}
          \label{eq:GravitationalIsing}
\hat{H}(t)=\int dx \ \hat{\bar{\psi}}\left[-\mathrm{i}\gamma^{\mu}\partial_{mu}+m_{\text{eff}}(t)\right]\hat{\psi}(x)
\end{align}
where the effective time dependent mass depends on the conformal factor of the metric as in Ref.~\cite{koke2016dirac}
\begin{align}
          \label{eq:Metric}
ds^2=\Omega(x,t)(dt^2-dx^2)
\ .
\end{align}
A similar Dirac Hamiltonian can be also obtained for the Friedmann-Robertson-Walker (FRW) metric with a scale factor that can be obtained by solving the Einstein's equations.

The common characteristic of these two examples is that a classical field satisfying equations of motion can drive the dynamics of a quantum field. It is important to note that the previous two examples are rather general non-relativistic and relativistic field theories so they are low-energy descriptions of a plethora of phenomena in diverse systems.

\section{Electronic structure driven by the nuclear motion
\label{AppendixD}}
To illustrate the generality of our approach, let us start by considering a generic example of quantum fields in second quantization that are driven by classical fields. One of them is inspired by the problem of electronic structure in quantum chemistry under the Born-Oppenheimer approximation~\cite{Born1927}.

The electronic structure problem can be written in terms of a non-relativistic field theory governed by the Hamiltonian~\cite{marques2004time}
\begin{align}
          \label{eq:FermionicHamiltonian}
\hat{H}(t)&=\int dx \ \hat{\psi}^{\dagger}(x)\left\{-\frac{\hbar^2\nabla^2}{2m}+V_{\text{ext}}[x,\boldsymbol{R}(t)]\right\}\hat{\psi}(x)
\nonumber \\
&+\int \int dx \ dx' \ \hat{\psi}^{\dagger}(x)\hat{\psi} ^{\dagger}(x')U(x,x')\hat{\psi}(x')\hat{\psi}(x)
\ ,
\end{align}
where $\hat{\psi}(x)$ denotes a fermionic quantum field satisfying the anti-commutation relation $\{\hat{\psi}(x),\hat{\psi}^{\dagger}(x') \}=\delta(x-x')$. Here the external potential being the interaction energy between the electrons and the nuclei
\begin{align}
          \label{eq:ExternalPotential}
V_{\text{ext}}[x,\boldsymbol{R}(t)] =-\sum^M_{\nu=1}\frac{Z_{\nu}}{|x-\boldsymbol{R}(t)|}
\end{align}
that is modulated by the positions $\boldsymbol{R}(t)=[\boldsymbol{R}_1(t),\dots,\boldsymbol{R}_M(t)]$ of the $M$ nuclei in the molecule that follow a classical trajectory that is a solution a classical equation of motion
\begin{align}
          \label{eq:ClassicalNuclei}
m_i\frac{d^2}{dt^2}\boldsymbol{R}_i(t)=-\nabla_{\boldsymbol{R}_i}U[\boldsymbol{R}(t)]
\ ,
\end{align}
where $U[\boldsymbol{R}(t)]$ is the potential energy. This is extremely important in the theoretical investigation of scattering  and chemical reactions. Similar models can be considered by problems of interacting bosons in external potentials.

One can decompose the fermionic field operator in terms of molecular or atomic orbitals $\hat{\psi}(x)=\sum_i \phi_i[x,\boldsymbol{R}(t)]\hat{a}_j$ as
 and by replacing this in Eq.~\eqref{eq:FermionicHamiltonian}, we obtain the Hamiltonian

\begin{align}
        \label{eq:HamiltonianSecondQuantization}
    \hat{H}[\boldsymbol{R}(t)]&=\sum_{i,j}h_{i,j}[\boldsymbol{R}(t)]\hat{a}^{\dagger}_{i}\hat{a}_j+\sum_{i,j}V_{i,j,k,l}[\boldsymbol{R}(t)]\hat{a}^{\dagger}_i\hat{a}^{\dagger}_j\hat{a}_k\hat{a}_l
    \ .
\end{align}
Depending on the chosen basis, the atomic or molecular orbitals may depend on the nuclear coordinates and for this reason, the nuclear motion induce a time dependence in the one- and two-particle integrals $h_{i,j}[\boldsymbol{R}(t)]$ and $V_{i,j,k,l}[\boldsymbol{R}(t)]$, respectively.

To simulate the electronic structure problem in a quantum computer, it is convenient to organize the fermionic operators $\hat{a}_j$ in a one-dimensional array and to use the Eq.~\eqref{eq:JordanWigner} to map the fermionic operators to qubits. By using this, we obtain the spin Hamiltonian
\begin{align}
    \label{eq:PauliWords}
\hat{H}_{\text{P}}(\boldsymbol{R})&=\sum_{\boldsymbol{J}}B_{\boldsymbol{J}}[\boldsymbol{R}(t)]\hat{P}_{\boldsymbol{J}}
          \ ,
\end{align}
where we define the Pauli words $\hat{P}_{\boldsymbol{J}}=\bigotimes^{2K}_j {\sigma}^{\boldsymbol{J}}_j$, which are elements of the Pauli group. The operator $\sigma^{\boldsymbol{J}}_j$ is one of the Pauli operators $\sigma^x_j,\sigma^y_j,\sigma^z_j$ and the identity $\hat{1}$ acting on the $j$-th qubit.  Note that again, there is parametric dependence of the coefficients $B_{\boldsymbol{J}}[\boldsymbol{R}(t)]$ on the nuclear trajectories $\boldsymbol{R}(t)$. The nuclei act as an external drive for the spin system.

\end{document}